# Insulator to metal transition of $WO_3$ epitaxial films induced by electrochemical Li-ion intercalation

K. Yoshimatsu[1,a)], T. Soma[1], and A. Ohtomo[1,2]

[1]*Department of Chemical Science and Engineering, Tokyo Institute of Technology, 2-12-1 Ookayama, Meguro-ku, Tokyo 152-8552, Japan*

[2]*Materials Research Center for Element Strategy (MCES), Tokyo Institute of Technology, Yokohama 226-8503, Japan*

a) electronic mail: k-yoshi@apc.titech.ac.jp

We investigated systematic evolutions of structural and electronic properties of $Li_xWO_3$ films, induced by Li-ion electrochemical reactions. Chronoamperometric Li-ion intercalation could control the amount of Li content up to $x \sim 0.5$. The resistivity abruptly decreased with increasing $x$ and the films underwent an insulator to metal transition (IMT) within a range of $0.2 < x < 0.24$, which was consistent with IMT of cubic $Na_xWO_3$. The X-ray diffraction analyses revealed the coexistence of tetragonal and cubic phases across IMT, suggesting that the alkaline-ion content was a primary factor for metallic conductivity in the $ReO_3$-type $WO_3$ system.

A 5$d$ transition-metal oxide of tungsten trioxide (WO$_3$) has been paid attention from the viewpoint of electrochromic application, *e.g.* smart windows.[1,2] WO$_3$ with a distorted ReO$_3$-type structure can be regarded as an *A*-site deficient perovskite, being capable of a host material for intercalations of alkaline and alkaline-earth metals.[3–7] Smaller H and Li atoms are easily intercalated to the amorphous or polycrystalline films, and the resultant doping electrons color WO$_3$. Because coloration of $A_x$WO$_3$ is attributed to increasing number of electrons in the W 5$d$ states, it will be associated with an electronic phase transition such as insulator to metal transition (IMT). In fact, IMT was reported for H$_x$WO$_3$ at $x \sim 0.17$[8] and $x = 0.32$.[9] On the other hand, Berggren *et al*. reported that the Li-ion intercalated Li$_x$WO$_3$ did not show IMT with Li content up to $x \sim 0.5$.[5] In these studies, amorphous and/or polycrystalline WO$_3$ were utilized for transport measurements. Therefore, it is still controversial whether crystalline quality and/or structure of the host affect critical content for IMT.

Electrochromic intercalation is one of the powerful approaches for systematic investigation of physical properties involving structural phase transition. A series of data can be taken with a single sample by repeating intercalation and measurement processes. Nevertheless, there have been no reports of the electronic and structural analyses for $A_x$WO$_3$ ($A$ = H and Li) along the intercalation.

In this Letter, we report structural and electronic properties of Li-ion intercalated epitaxial $WO_3$ films. Systematic Li-ion intercalation to $WO_3$ films enabled us to access the evolution of physical properties as a function of Li content. We revealed drastic decrease in resistivity and the occurrence of IMT with increasing Li content. The emergence of metallicity was discussed based on the results of resistivity and X-ray diffraction (XRD). The critical content for IMT in $Li_xWO_3$ was found to be close to that in $Na_xWO_3$, suggesting that amount of W $5d$ electrons is a primary factor for metallic conductivity in the $ReO_3$-type $WO_3$ systems.

$WO_3$ films were grown on $LaAlO_3$ (100) substrates by using pulsed-laser deposition (PLD) technique in an ultrahigh vacuum chamber. KrF excimer laser pulses (25 Hz, 1.0 J/cm$^2$) were focused on a polycrystal $WO_3$ ceramic tablet. Substrate temperature was set 600ºC and oxygen pressure in the PLD chamber was fixed at 100 mTorr with continuous flow of pure oxygen gas (6N purity). Film thickness (~30 nm) was evaluated by glazing incidence X-ray reflectometry. The flat surface with rms roughness of the as-grown films (~1.0 nm) was verified by using atomic force microscopy.

The Li-ion electrochemical reactions were performed with a standard three-electrode system. The $WO_3$ films, $LiCoO_2$/Al foil (AA Portable Power

Corporation), and Ag wire (Nilaco, 4N purity) were used for working, counter, and quasi-reference electrodes, respectively. A liquid electrolyte (1 mol/L) was prepared by mixing Li-ion salt of anhydrous $LiClO_4$ (purity > 99%) and organic solvent of anhydrous propylene carbonate (PC) (purity > 98%) in a glovebox with Ar atmosphere. It was heated at ~50ºC to remove moisture before immersing the electrodes. Cyclic voltammetry and chronoamperometry were performed with VersaSTAT4 (ALS Co., Ltd) at RT.

Temperature-dependent resistivity ($\rho$-$T$) curves were measured by a four-probe method using physical properties measurement system (Quantum design, PPMS). Al wires with a diameter of 25 μm were directly welded to the film surface. The crystal structures of the $WO_3$ films before and after Li-ion electrochemical reactions were characterized by a XRD apparatus with Cu K$\alpha_1$ radiation.

Figure 1 shows typical cyclic voltammogram of a $WO_3|LiClO_4:PC|LiCoO_2$/Al electrochemical cell. The reductive current was clearly observed in a potential range from −2.0 to 0 V vs. Ag/AgClO$_4$ quasi-reference electrode. The observed cyclic voltammogram is similar to those reported previously, except for anodic peak separation.[3,7,10] This difference is due to relatively higher voltage-sweep rate (100 mV/s) used in this study.

The amount of intercalated Li ions was restricted with a chronoamperometric method. Figure 2 (a) shows typical chronoamperogram of the electrochemical cell measured by applying a potential of −1.0 V vs. Ag/AgClO$_4$. The observed cathodic current and rapid decay to a background level (−1 μA after ~5 s) suggests that Li ions and an equivalent amount of electrons are simultaneously introduced to the WO$_3$ film. The amount of the intercalated Li ions, *i.e.*, the total charge during Li-ion electrochemical reactions, was calculated from the hatched area in Fig. 2(a) using the following equation,

$$x = \frac{\int_0^\infty I\, dt}{F \times n_{WO3}},$$

where $F$ and $n_{WO3}$ are Faraday constant and a molar amount of WO$_3$, respectively. As shown in inset of Fig. 2(a), $x$ increased monotonically with increasing applied potential. We tested a number of samples having different thickness and surface area and concluded that all current was faradaic and a statistical error in $x$ was no greater than 5 %. Li content as high as $x \sim 0.5$ could be achieved by applying a potential of −2.5 V vs. Ag/AgClO$_4$. Application of more negative potential caused degradation of crystal structure as discussed later.

The evolution of electronic properties of Li$_x$WO$_3$ films was investigated for resistivity at 300 K ($\rho_{300\,K}$). Figure 2 (b) shows plot of $\rho_{300\,K}$ as a function of $x$. The electrochemical cell as illustrated in inset of Fig. 2 (b) was used for the repetition of Li-ion intercalation and resistivity measurements. For the as-grown film, $\rho_{300\,K}$ was ~$10^2$ $\Omega$ cm. The resistivity drastically decreased with increasing $x$; it became ~$5\times10^{-4}$ $\Omega$ cm for the sample with $x \sim 0.10$ and gradually decreased for those with $x > 0.1$.

The electronic phase of Li$_x$WO$_3$ films was identified from $\rho$-$T$ curves. For this measurement, the Li-ion intercalation reaction was carried out at 300 K in PPMS using the setup shown in inset of Fig. 2(b). Then, the film was cooled down to 230 K (a freezing point of LiClO$_4$:PC electrolyte), and meanwhile an open-circuit potential was applied to prevent self-discharge (Li-ion deintercalation). After freezing the electrolyte, we started $\rho$-$T$ measurement without applying potential. After warming up to 230 K, we applied an open-circuit potential again to return to RT. As shown in Fig. 3(a), the sample remained insulating during several repeats, while the resistivity systematically decreased in the entire range of temperature.

The $\rho$-$T$ curves in the vicinity of IMT are shown in Fig. 3(b). The insulating behavior was still seen at $x \leq 0.20$. Meanwhile, the sample with $x \geq 0.24$ turned to be metallic, exhibiting a slight upturn at low temperature ($T < 50$ K). Therefore, IMT took place in a range of $0.20 < x < 0.24$. The metallic behavior remained down to 10 K for the sample with $x \geq 0.28$.

Structural transitions of the $Li_xWO_3$ films were studied by XRD measurements. Figure 4 shows out-of-plane XRD patterns around $LaAlO_3$ 200 reflection for the $Li_xWO_3$ films with various $x$ (applied potentials). The reflections from the films were detected at the higher angle side of the $LaAlO_3$ 200 reflection. As for the as-grown $WO_3$ film, it was detected at $2\theta = 24.18°$, corresponding to $d = 3.680$ Å. Because $\gamma$-phase is most stable at RT, the observed reflection can be assigned to $\gamma$-$WO_3$ 200 ($d = 3.653$ Å).[11,12] The rocking curve of the $\gamma$-$WO_3$ 200 reflection indicated FWHM of ~ 0.1° (not shown). Slight deviation from the bulk value is due to compressive epitaxial strain from the substrate as discussed later.

The peak position remained intact even after applying negative potentials of −0.25 and −0.50 V vs. $Ag/AgClO_4$. When applying a potential of −1.5 V ($x = 0.13$), the film reflection split into two peaks, which were located at $2\theta = 23.85°$ ($d = 3.727$ Å) and the same angle as that of as-grown $WO_3$, respectively. After applying a potential of −2.5 V ($x = 0.51$), the original peak vanished and only lower-angle peak was detected. The film reflections were only found at higher angle side of $LaAlO_3$ ($h$00) reflections ($h = 1, 2$) throughout the experiments, suggesting a complete structural phase transition to cubic. We note that the epitaxial structure was preserved up to $x \sim 0.5$. By applying more negative potential (−3.0 V vs. Ag/AgClO$_4$), the film reflection became broader and very weak due to degradation of crystallinity. Therefore, we conclude that the samples with Li content up to $x \sim 0.5$ have high crystallinity.

Let us discuss the structural and electronic properties of Li$_x$WO$_3$ films as a function of $x$. The as-grown WO$_3$ film took $\gamma$-phase (monoclinic structure: $a = 7.306$ Å, $b = 7.540$ Å, $c = 7.692$ Å, $\beta = 90.88°$). The $a$-axis oriented WO$_3$ film grew on pseudocubic LaAlO$_3$ (100) substrates ($2a_{pc} = 7.58$ Å) because of relatively small lattice mismatches along the $b$- and $c$-axes (−0.53 % and 1.48 %, respectively). In this case, compressive strain along the $c$-axis overcomes tensile strain along the $b$-axis, giving rise to expansion of the $a$-axis lattice constant. On the other hand, Li$_x$WO$_3$ films with $0.01 \leq x \leq 0.082$ should have been partially transformed to tetragonal phase according to a bulk phase diagram.[3] Nevertheless, only single peak was detected for the Li$_x$WO$_3$ film with $x = 0.026$. We presume that the out-of-plane $d$ values between the strained monoclinic and relaxed tetragonal crystals are too close to be discriminated each other. In fact, the observed out-of-plane lattice constant was almost identical to that of bulk Li$_{0.10}$WO$_3$.[3] Both of the tetragonal and cubic phases were detected for the Li$_{0.13}$WO$_3$ film, while only cubic phase appeared in the Li$_x$WO$_3$ film ($x \geq 0.36$). Taken together, all the XRD patterns are consistent with bulk phase diagram.

According to the $\rho$-$T$ curves, IMT occurred at $0.2 < x < 0.24$ in our Li$_x$WO$_3$ film. As shown in Fig. 5, the tetragonal and cubic phases coexist across MIT, which means no direct correlation between the structural and electronic phase transitions. Previous studies on other doped WO$_3$ systems help us to discuss the electronic phase transition in Li$_x$WO$_3$. The crystal structures of $A_x$WO$_3$ ($A$: H and alkaline metals) are classified into the ReO$_3$ and hexagonal types.[13] The former (latter) are thermodynamically stabilized when the smaller (larger) $A$-ions such as H$^+$, Li$^+$, and Na$^+$ (K+, Rb+, and Cs+)[14–17] are doped into the $A$-sites. Therefore, Na$_x$WO$_3$ is the most relevant to Li$_x$WO$_3$ for comparing electronic structures as a function of $x$.[13, 18–20] Na$_x$WO$_3$ is known to show IMT at the critical content of $0.23 < x < 0.29$[20], very close to what we observed for Li$_x$WO$_3$. Unlike Li$_x$WO$_3$, however, the metallic Na$_x$WO$_3$ has a cubic structure. The previous theoretical study on Na$_x$WO$_3$ explains that IMT is driven by electronic factors (W $5d$ electrons).[21] If IMT in Li$_x$WO$_3$ is driven by similar electronic factors, the large amount of doped electrons will hide the small difference in the crystal structures. In order to obtain further evidence, quantitative H-intercalation and/or electrostatic doping[22] to the epitaxial WO$_3$ films are better attempts for future experiments.

In summary, we studied evolution of the crystal structure and electronic phase for Li$_x$WO$_3$ films.  Using an electrochemical cell with chronoamperometric method, large amount of Li ions (up to $x \sim 0.5$) could be introduced into epitaxial WO$_3$ films.  A series of resistivity measurements revealed the systematic decrease in resistivity and IMT for Li$_x$WO$_3$ films with $0.2 < x < 0.24$, around which IMT also occurred for Na$_x$WO$_3$ with cubic structure.  The Li$_x$WO$_3$ films in the vicinity of IMT exhibited a mixed structure of the tetragonal and cubic phases. These results suggest that IMT is governed by not crystal structures but critical content of $A$-ions for the ReO$_3$-type WO$_3$ systems.

**Acknowledgements**     The authors thank T. Takao, M. Oishi and M. Nagaoka for assistance regarding preparation of electrolyte.  This work was partly supported by MEXT Elements Strategy Initiative to Form Core Research Center and a Grant-in-Aid for Scientific Research (Nos. 15H03881 and 16H05983) from the Japan Society for the Promotion of Science Foundation.

**Figure Captions**

**Fig. 1.** Cyclic voltammogram of a WO$_3$|LiClO$_4$:PC|LiCoO$_2$/Al electrochemical cell measured at room temperature with a voltage sweep rate of 100 mV/s.

**Fig. 2.** (a) Chronoamperogram of Li$_x$WO$_3$ films by applying a potential of −1.0 V vs. Ag/AgClO$_4$. The inset shows the amount of the intercalated Li ions as a function of applied potential. (b) Resistivity at 300 K ($\rho_{300\,K}$) as a function of Li content. The inset shows schematic illustration of the Li-ion electrochemical cell used in this study.

**Fig. 3.** Temperature dependence of resistivity for Li$_x$WO$_3$ films with (a) smaller and (b) larger Li content.

**Fig. 4.** Out-of-plane XRD patterns of the as-grown WO$_3$ and Li-ion intercalated Li$_x$WO$_3$ films under various potentials (Li content).

**Fig. 5.** Structural and electronic phase diagram of Li$_x$WO$_3$. The $d$ values are plotted against Li content $x$ for the Li$_x$WO$_3$ films and bulk samples.[3] *M*, *T*, and *C* denote monoclinic, tetragonal and cubic structures, respectively.

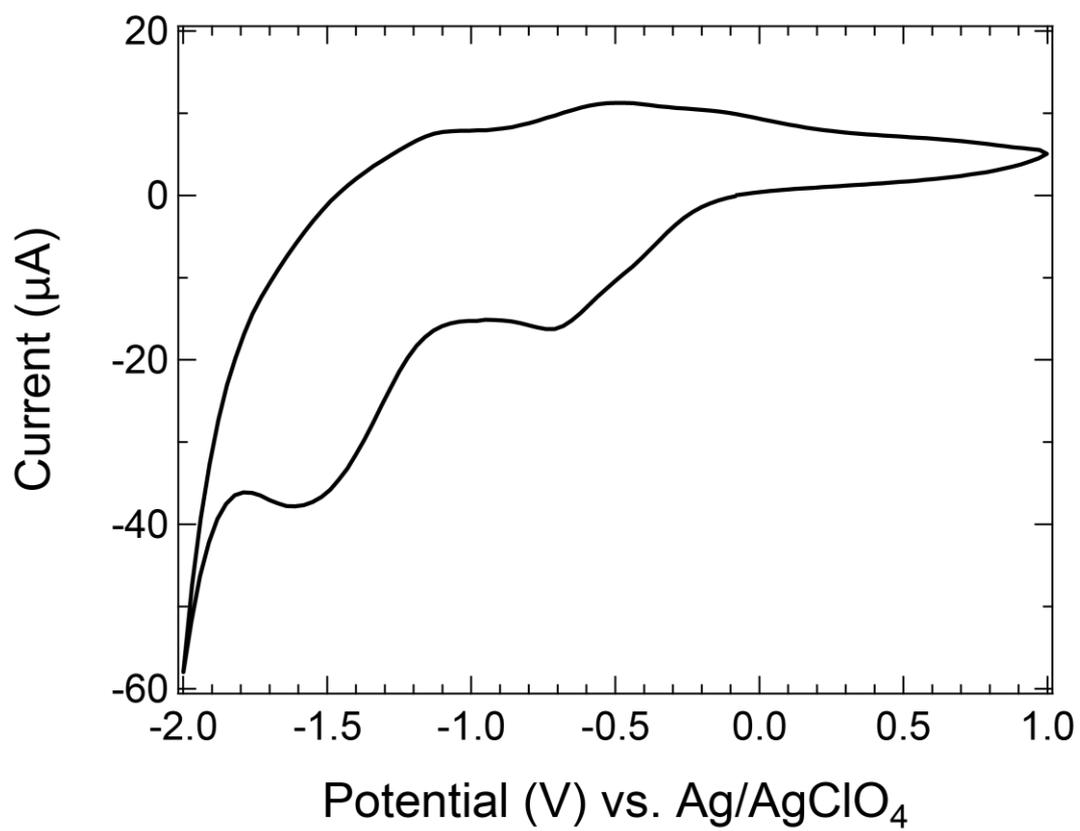

Figure 1 K. Yoshimatsu *et al*.

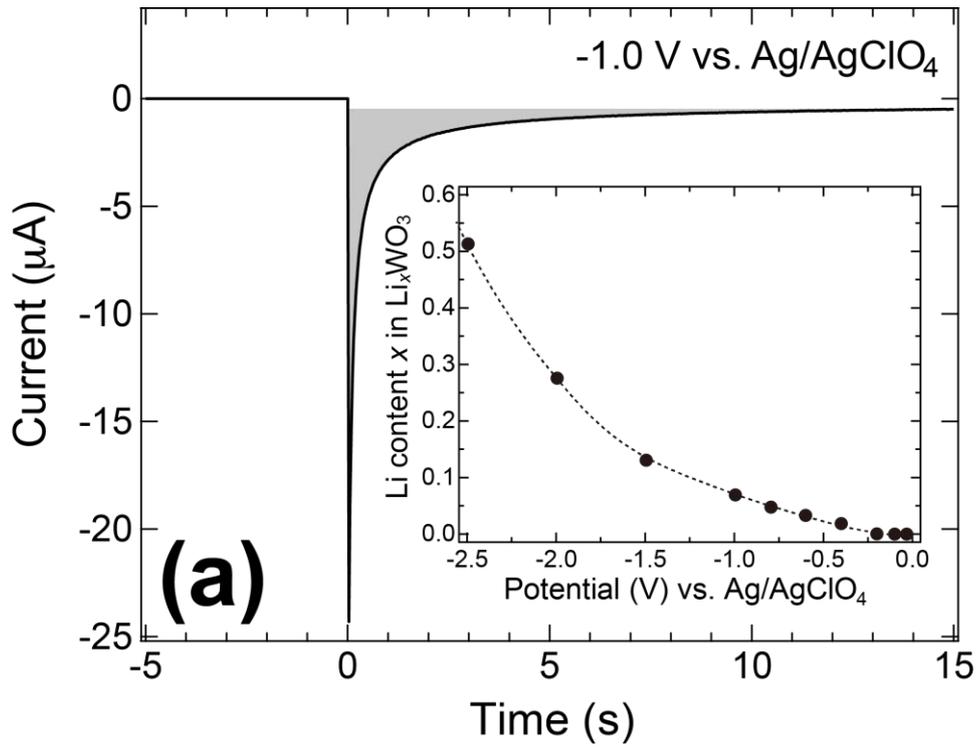

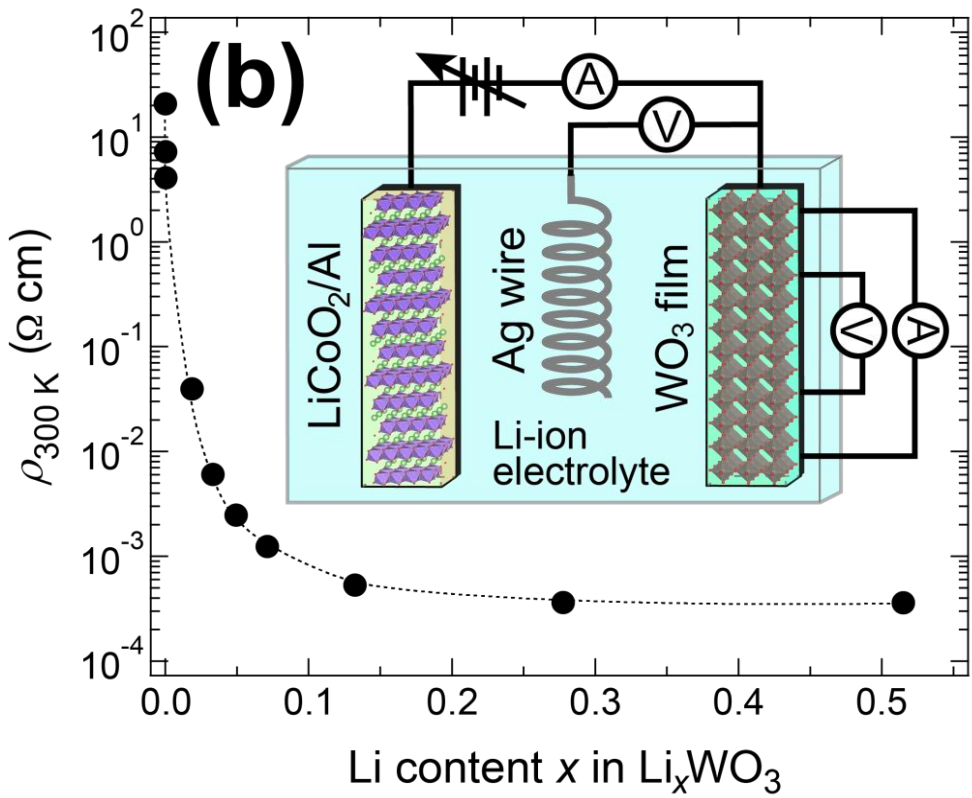

Figure 2 K. Yoshimatsu *et al*.

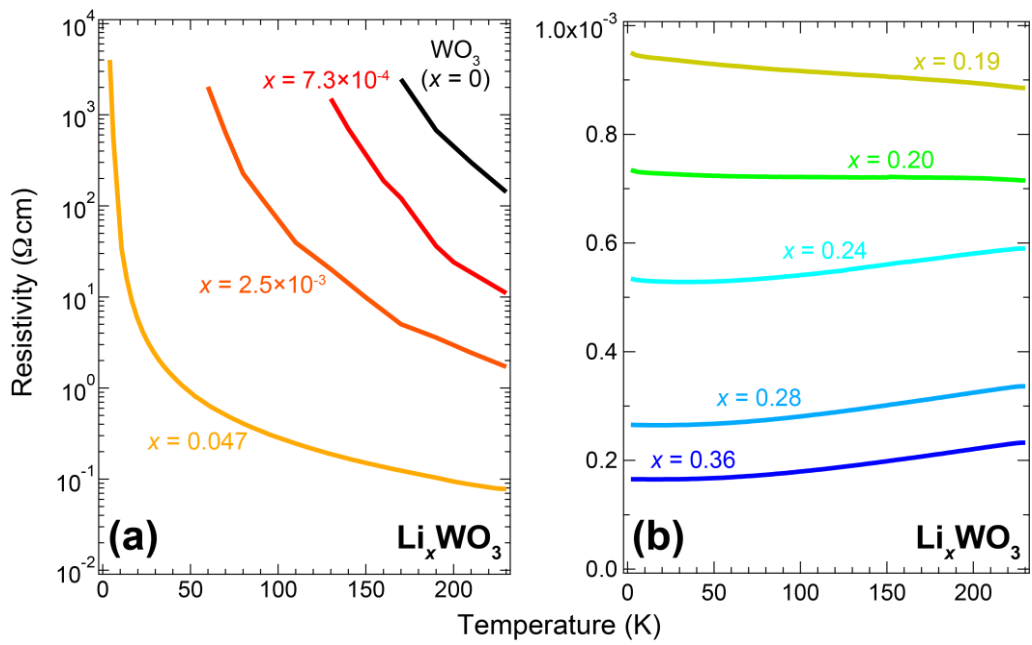

Figure 3 K. Yoshimatsu *et al*.

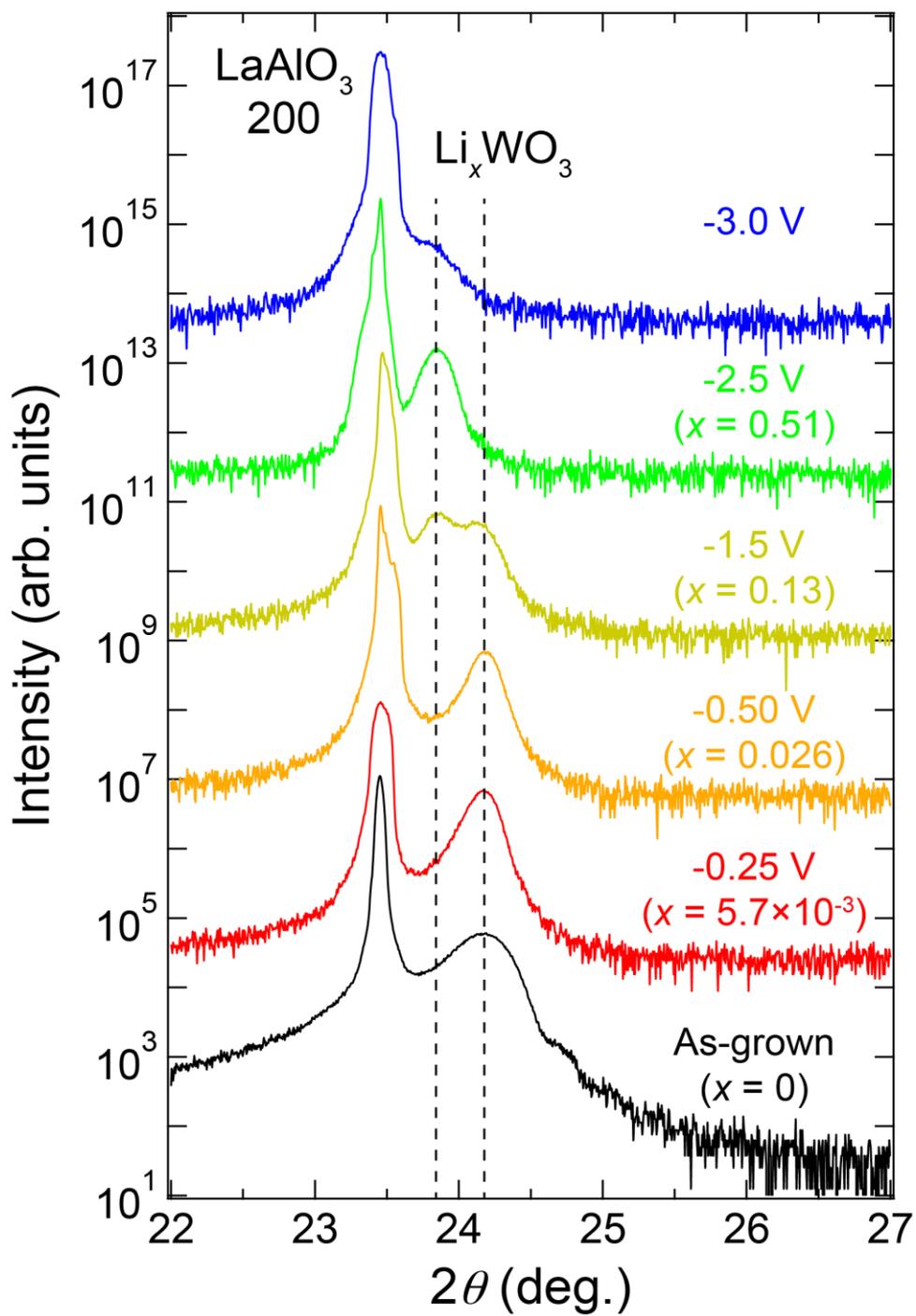

Figure 4 K. Yoshimatsu *et al*.

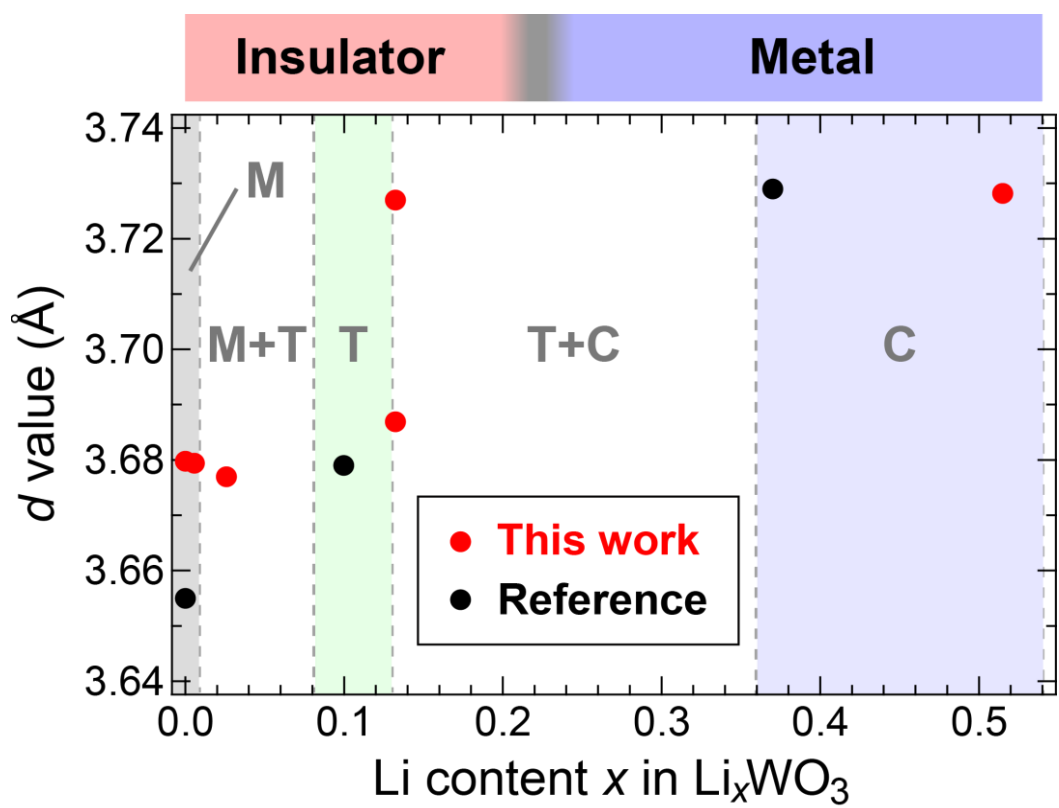

Figure 5 K. Yoshimatsu *et al*.